\documentstyle[prd,eqsecnum,aps,epsf]{revtex}

\def\be{\begin{equation}}
\def\ee{\end{equation}}
\def\ba{\begin{eqnarray}}
\def\ea{\end{eqnarray}}

\newcommand{\beqa}{\begin{eqnarray}}
\newcommand{\eeqa}{\end{eqnarray}}
\newcommand{\bea}{\begin{eqnarray}}
\newcommand{\eea}{\end{eqnarray}}

\newcommand{\hp}{\hat{\phi}}
\newcommand{\hV}{\hat{V}}

\newcommand{\singlefig}[2]{
\begin{center}
\begin{minipage}{#1}
\epsfxsize=#1
\epsffile{#2}
\end{minipage}
\end{center}}
\newenvironment{figcaption}[2]{
 \vspace{0.3cm}
 \refstepcounter{figure}
 \label{#1}
 \begin{center}
 \begin{minipage}{#2}
 \begingroup \small FIG. \thefigure: }{
 \endgroup
 \end{minipage}
 \end{center}}
\def\beq{\begin{equation}}
\def\eeq{\end{equation}}

\newcommand{\square}{\kern1pt\vbox{\hrule height
1.2pt\hbox{\vrule width 1.2pt\hskip 3pt
   \vbox{\vskip 6pt}\hskip 3pt\vrule width 0.6pt}\hrule
height 0.6pt}\kern1pt}

\begin{document}

\title{Density perturbations in generalized Einstein scenarios and 
constraints on \\
nonminimal couplings from the Cosmic Microwave Background}
\author{Shinji Tsujikawa$^1$ and Burin Gumjudpai$^2$}
\address{$^1$Institute of Cosmology and Gravitation, 
University of Portsmouth, Mercantile House, 
Portsmouth PO1 2EG, \\
United Kingdom \\[.3em]} 
\address{$^2$Fundamental Physics \& Cosmology Research Unit, 
The Tah Poe Group of Theoretical Physics (TPTP), \\
Department of Physics, Naresuan University, Phitsanulok, 
Thailand 65000 \\[.3em]} 

\date{\today} 
\maketitle
\begin{abstract}
We study cosmological perturbations in generalized Einstein scenarios 
and show the equivalence of inflationary observables both in 
the Jordan frame and the Einstein frame.
In particular the consistency relation relating the tensor-to-scalar
ratio with the spectral index of tensor perturbations 
coincides with the one in Einstein gravity, which leads to 
the same likelihood results in terms of inflationary observables. 
We apply this formalism to nonminimally coupled chaotic inflationary 
scenarios with potential $V=c\phi^p$ and place constraints 
on the strength of nonminimal couplings   
using a compilation of latest observational data.
In the case of the quadratic potential ($p=2$), the nonminimal coupling 
is constrained to be $\xi>-7.0 \times 10^{-3}$ for negative 
$\xi$ from the $1\sigma$
observational contour bound. Although the quartic potential ($p=4$)
is under a strong observational pressure for $\xi=0$, this property
is relaxed by taking into account negative nonminimal couplings.
We find that inflationary observables are within the $1\sigma$ 
contour bound as long as $\xi<-1.7 \times 10^{-3}$.  
We also show that the $p \ge 6$ cases are disfavoured 
even in the presence of nonminimal couplings.
\end{abstract}

\vskip 1pc 

\baselineskip = 12pt

\section{Introduction}                            

The inflationary paradigm has been the backbone of high-energy 
cosmology over the past 20 years \cite{review}.
The striking feature of the inflationary cosmology is that it 
predicts nearly scale-invariant, gaussian, adiabatic density perturbations
in its simplest form. This prediction shows an excellent agreement  
with all existing and accumulated data within
observational errors. In particular the recent measurement of the Wilkinson Microwave Anisotropy Probe 
(WMAP) \cite{Spergel} provided the high-precision  
dataset from which inflationary models can be seriously constrained 
\cite{Bridle,Peiris,Barger,Kinney,Leach,Tegmark}. 

The prediction of inflationary observables exhibits some difference 
depending on the models of inflation. It is important to pick up 
this slight difference in order to discriminate between a host 
of inflationary scenarios. Conventionally inflationary models
can be classified in three classes \cite{Kolb}--``large-field", ``small-field", and 
``hybrid" models-- depending upon the shape of the inflaton potential.
The large field model is characterized by the potential
\begin{eqnarray}
V(\phi)=c \phi^p\,,
\label{potential}
\end{eqnarray}
which includes only one free parameter for a fixed value of $p$.
Therefore it is not generally difficult to place strong constraints
on the potential (\ref{potential}) compared to small-field and hybrid 
models that involve additional model parameters.
In fact it was recently shown in 
Refs.\,\cite{Peiris,Barger,Kinney,Leach,Tegmark} that 
the quartic potential ($p=4$) is under a strong observational pressure 
due to the deviation of a scale-invariant spectrum in addition to a high 
tensor-to-scalar ratio. This situation does not change much 
in the context of the Randall-Sundrum II braneworld 
scenario \cite{LS,TL}.
Note that the quadratic potential ($p=2$) is within observational 
contour bounds due to the flatness of the potential relative to 
the quartic case.

When we try to construct models of inflation based on particle physics,
we do not need to restrict ourselves to the standard Einstein gravity.
For example, low-energy effective string theory gives rise to a
coupling between the scalar curvature and the dilaton field, which 
leads to an inflationary solution in the string frame \cite{PBB}. 
The Jordan-Brans-Dicke (JBD)-like theories can be viewed as the low-energy 
limit of superstring theory if the Brans-Dicke scalar plays the role 
of the dilaton \cite{FT85,C85}.
In this sense the proposal of the extended inflation scenario \cite{extended} 
stimulated a further study of more generic classes of inflation models in 
non-Einstein theories \cite{soft,Lin90}, in spite of the fact that the first version of 
the extended inflation resulted in failure due to the graceful exit 
problem \cite{presc}.

From the viewpoint of quantum field theory in curved spacetime, 
nonminimal couplings naturally arise due to their own 
nontrivial renormalization group flows.
The ultraviolet fixed point of these flows is often divergent,
implying that nonminimal couplings can be important 
in the early universe.
In this respect Futamase and Maeda \cite{FM} studied the effect of 
nonminimal couplings on the dynamics of chaotic inflation and 
estimated the strength of the coupling $\xi$ from the requirement 
of a sufficient inflation. While $|\xi|$ is required to be much smaller than 
unity in the quadratic potential, such a constraint is absent in the 
quadratic potential for negative $\xi$. 
Fakir and Unruh \cite{FU} showed that the fine-tuning 
problem of the self-coupling $c$ in Eq.~(\ref{potential}) is relaxed
for large negative values of $\xi$ by evaluating 
the amplitude of scalar metric perturbations.
A number of authors further investigated the spectra of scalar and
tensor perturbations generated in inflation 
\cite{Salopek,Makino,Salopek2,Kaiser,KF0,KF,SY,massive} and 
particle productions in 
reheating \cite{preheating}\footnote{The presence of negative 
nonminimal couplings 
also leads to the strong variation of curvature perturbations in the 
context of multi-field inflation. See Refs.~\cite{multi} for 
details.}.
See e.g., Refs.~\cite{Gun} for other interesting aspects
of nonminimally coupled scalar fields.

In this work we do not restrict ourselves to the Fakir and Unruh scenario,
but will place constraints on the strength of nonminimal coupling 
for the general potential (\ref{potential}) by using a compilation of 
recent observational datasets. We shall provide a general formalism 
for scalar and tensor perturbations in generalized Einstein theories
including dilaton gravity, JBD theory and a nonminimally coupled 
scalar field. This analysis explicitly shows the equivalence of inflationary 
observables in both the Jordan frame and the Einstein frame, 
which results in the fact that a separate likelihood analysis 
in terms of observational quantities is not needed compared to 
the Einstein gravity. 
Making use of the 2-dimensional observational constraints of
the scalar spectral index $n_{\rm S}$ and the tensor-to-scalar ratio $R$,
we shall carry out a detailed analysis about the
constraint on the nonminimally coupled inflaton field 
for the potential (\ref{potential}).
This provides us an interesting possibility to place strong constraints
on the strength of nonminimal couplings from the recent 
high-precision observations.
In fact we  will show that even small nonminimal couplings 
with $|\xi| \ll 1$ can alter the standard prediction 
of the Einstein gravity.

\section{General formalism for perturbation spectra}                            

We start with a generalized action \cite{Hwang}
\begin{eqnarray}
{\cal S}=\int d^4 x \sqrt{-g} \left[ \frac12 F(\phi)R
-\frac12 \omega(\phi) (\nabla \phi)^2-V(\phi)\right]\,,
\label{lag}
\end{eqnarray}
where $R$ is the Ricci scalar.
$F(\phi)$, $\omega(\phi)$ and $V(\phi)$ are 
general functions of a scalar field $\phi$.  The action (\ref{lag}) 
includes a variety of gravity theories such as the Einstein gravity,
scalar tensor theories and low-energy effective string theories.
For example, we have $F(\phi)=1/\kappa^2$ and $\omega(\phi)=1$ in 
the Einstein gravity, $F(\phi)=(1-\xi \kappa^2 \phi^2)/\kappa^2$
and $\omega(\phi)=1$ for a nonminimally coupled scalar field, 
$F(\phi)=e^{-\phi}$ and $\omega(\phi)=-e^{-\phi}$ for 
low-energy effective string theories.
Hereafter we basically use the unit 
$\kappa^2 \equiv 8\pi/m_{\rm pl}^2=1$, but 
we restore the Planck mass $m_{\rm pl}$ when it is needed.

In a flat Friedmann-Lemaitre-Robertson-Walker (FLRW) background 
with a scale factor $a$, the background equations are 
\begin{eqnarray}
\label{back2}
& & H^2 \equiv \left(\frac{\dot{a}}{a}\right)^2=\frac{1}{6F} 
\left(\omega \dot{\phi}^2+ 2V-6H\dot{F}\right)\,,~~~\\
& & \dot{H}=\frac{1}{2F} \left( -\omega\dot{\phi}^2+H\dot{F}
-\ddot{F}\right)\,, \\
& &\ddot{\phi}+3H\dot{\phi}+\frac{1}{2\omega}
\left(\omega_\phi \dot{\phi}^2-F_\phi R+2V_\phi\right)=0\,.
\label{back}
\end{eqnarray}
where $H$ is the Hubble expansion rate and a dot denotes the 
derivative with respect to a cosmic time $t$.

\subsection{Perturbations in Jordan frame}  

We consider a general perturbed metric for scalar perturbations 
\begin{eqnarray}
ds^2 = -(1+2A)dt^2 +2a(t)B_{,i}dx^idt
+a^2(t)\left[(1+2\varphi)\delta_{ij} 
+2E_{,i,j}\right]dx^i dx^j\,,
\label{pmetric}
\end{eqnarray}
where a comma denotes a flat-space coordinate 
derivative. 
It is convenient to introduce
comoving curvature perturbations, ${\cal R}$, defined by 
\begin{eqnarray}
{\cal R}=\varphi-\frac{H}{\dot{\phi}}\delta \phi\,,
\label{calR}
\end{eqnarray}
where $\delta \phi$ is the perturbation of the field $\phi$.
The equation of motion for the lagrangian (\ref{lag}) was derived 
in Refs.\,\cite{Hwang} and is simply given by 
\begin{eqnarray}
\frac{1}{a^3Q_{\rm S}} \left(a^3 Q_{\rm S} \dot{\cal R}\right)^{\bullet}
+\frac{k^2}{a^2}{\cal R}=0\,,~~~{\rm with}~~~
Q_{\rm S}=\frac{\omega \dot{\phi}^2+3\dot{F}^2/2F}
{(H+\dot{F}/2F)^2}\,,
\label{Req}
\end{eqnarray}
where $k$ is a comoving wavenumber.
Note that a similar form of equation is derived even in the presence 
of more complicated terms such as the Gauss-Bonnet term and 
the $(\nabla \phi)^4$ term \cite{Cartier}. 
If we neglect the contribution of the decaying mode, the curvature 
perturbation is conserved in the large-scale limit ($k \to 0$).
Introducing new variables, $z=a\sqrt{Q_{\rm S}}$ and $v=a{\cal R}$, the 
above equation reduces to 
\begin{eqnarray}
v''+\left(k^2-z''/z\right)v=0\,,
\label{veq}
\end{eqnarray}
where a prime denotes a derivative 
with respect to a conformal time $\eta=\int a^{-1}dt$.

The gravity term $z''/z$ can be written as 
\begin{eqnarray}
\frac{z''}{z}=(aH)^2 \left[(1+\delta_{\rm S})(2+\delta_{\rm S}
+\epsilon)
+\frac{\delta_{\rm S}'}{aH}\right]\,,
\label{gra}
\end{eqnarray}
where 
\begin{eqnarray}
\epsilon=\frac{\dot{H}}{H^2}\,,~~~
\delta_{\rm S}=\frac{\dot{Q}_{\rm S}}{2HQ_{\rm S}}\,.
\label{ep}
\end{eqnarray}
In the context of slow-roll inflation, it is a good approximation 
to neglect the variations of $\epsilon$ and $\delta_{\rm S}$.
Since $\eta=-1/[(1+\epsilon)aH]$ in this case, we have 
\begin{eqnarray}
\frac{z''}{z}=\frac{\gamma_{\rm S}}{\eta^2}\,,~~~
{\rm with}~~~
\gamma_{\rm S}=\frac{(1+\delta_{\rm S})
(2+\delta_{\rm S}+\epsilon)}
{(1+\epsilon)^2}\,.
\label{gra2}
\end{eqnarray}
Then the solution for Eq.~(\ref{veq}) is expressed 
by the combination of the Hankel functions: 
\begin{eqnarray}
v=(\sqrt{\pi |\eta|}/2)\left[ c_1 H_{\nu_{\rm S}}^{(1)}
(k|\eta|) +c_2 H_{\nu_{\rm S}}^{(2)}(k|\eta|) \right]\,,
\label{han}
\end{eqnarray}
where $\nu_{\rm S} \equiv \sqrt{\gamma_{\rm S}+1/4}$.   
We choose the coefficients to be $c_1=0$ and $c_2=1$, 
so that positive frequency solutions in a Minkowski
vacuum are recovered 
in an asymptotic past.
Making use of the relation $H_{\nu_{\rm S}}^{(2)} 
(k|\eta|) \to (i/\pi) 
\Gamma(\nu_{\rm S}) (k|\eta|/2)^{-\nu_{\rm S}}$ 
for long wavelength perturbations 
($ k \to 0$), one gets the spectrum of curvature perturbations,
${\cal P}_{\rm S} \equiv (k^3/2\pi^2)\left| {\cal R} \right|^2$, 
as 
\begin{eqnarray}
{\cal P}_{\rm S} = \frac{1}{Q_{\rm S}}
\left(\frac{H}{2\pi}\right)^2
\left(\frac{1}{aH|\eta|}\right)^2
\left(\frac{\Gamma(\nu_{\rm S})}{\Gamma(3/2)}\right)^2
\left(\frac{k|\eta|}{2}\right)^{3-2\nu_{\rm S}} \equiv 
A_{\rm S}^2 \left(\frac{k|\eta|}{2}\right)
^{3-2\nu_{\rm S}}\,. 
\label{PS}
\end{eqnarray}
When $\nu=0$, we have an additional ${\rm ln}\,(k|\eta|)$ 
factor \cite{Cartier}.

Then the spectral index, 
$n_{\rm S} \equiv 1+{\rm d}{\rm ln}{\cal P}_{\rm S}/{\rm d}{\rm ln}k$,
is given by 
\begin{eqnarray}
n_{\rm S}-1 =3-2\nu_{\rm S}=3-\sqrt{4\gamma_{\rm S}+1}\,,
\label{index}
\end{eqnarray}
where $\gamma_{\rm S}$ is given in Eq.~(\ref{gra2}).

Let us next consider the spectrum of tensor perturbations, $h_{ij}$.
Since $h_i^j$ satisfies the same form of equation as in Eq.~(\ref{Req})
with replacement $Q_{\rm S} \to Q_{\rm T}=F$, we get the power spectrum 
to be
\begin{eqnarray}
{\cal P}_{\rm T}=\frac{8}{Q_{\rm T}}\left(\frac{H}{2\pi}\right)^2
\left(\frac{1}{aH|\eta|}\right)^2
\left(\frac{\Gamma(\nu_{\rm T})}{\Gamma(3/2)}\right)^2
\left(\frac{k|\eta|}{2}\right)^{3-2\nu_{\rm T}} \equiv A_{\rm T}^2
\left(\frac{k|\eta|}{2}\right)^{3-2\nu_{\rm T}}\,,
\label{PT}
\end{eqnarray}
where 
\begin{eqnarray}
\nu_{\rm T}=\sqrt{\gamma_{\rm T}+1/4}\,,~~~
{\rm with}~~~
\gamma_{\rm T}=\frac{(1+\delta_{\rm T})
(2+\delta_{\rm T}+\epsilon)}
{(1+\epsilon)^2}~~~{\rm with}~~~
\delta_{\rm T}=\frac{\dot{Q}_{\rm T}}
{2HQ_{\rm T}}\,.
\end{eqnarray}
Note that we take into account the polarization states of 
gravitational waves. 
The spectral index of tensor perturbations
is given by
\begin{eqnarray}
n_{\rm T}=3-\sqrt{4\gamma_{\rm T}+1}\,.
\label{nT}
\end{eqnarray}
The difference of scalar and tensor perturbations comes from the fact that 
$Q_{\rm S}$ differs from $Q_{\rm T}$.

The tensor-to-scalar ratio is defined as
\begin{eqnarray}
R \equiv \frac{A_{\rm T}^2}{A_{\rm S}^2} =
8\frac{Q_{\rm S}}{Q_{\rm T}}
\left(
\frac{\Gamma(\nu_{\rm T})}{\Gamma(\nu_{\rm S})}
\right)^2=
8\frac{\omega \dot{\phi}^2+3\dot{F}^2/2F}
{F(H+\dot{F}/2F)^2}
\left(
\frac{\Gamma(\nu_{\rm T})}{\Gamma(\nu_{\rm S})}
\right)^2
\,.
\label{ratio}
\end{eqnarray}

\subsection{Einstein frame and the equivalence of perturbation spectra}  

The discussion in the previous subsection corresponds to 
the analysis in the Jordan frame. It is not obvious whether 
the same property holds in the Einstein frame.
Let us make the conformal transformation for 
the action (\ref{lag}) 
\begin{eqnarray}
\hat{g}_{\mu \nu}=\Omega g_{\mu \nu}\,,~~~
{\rm with}~~~\Omega=F\,.
\label{conformal}
\end{eqnarray}
Then we get the action in the Einstein 
frame \cite{Hwang2,Faraoni}
\begin{eqnarray}
{\cal S}_E=\int d^{4} \hat{x} \sqrt{-\hat{g}} 
\left[ \frac12 \hat{R}-\left\{ \frac34 \left(\frac{F'}{F}\right)^2+
\frac{\omega}{2F} \right\}(\hat{\nabla} \phi)^2-\hV (\phi) \right]\,,~~~
{\rm with}~~~\hV (\phi)=\frac{V(\phi)}{F^2}\,.
\label{Einstein}
\end{eqnarray}
If we introduce a new scalar field, $\hp$, so that 
\begin{eqnarray}
d\hp \equiv G(\phi)d\phi\,,~~~
{\rm with}~~~G(\phi)=
\sqrt{\frac32 \left(\frac{F'}{F}\right)^2+
\frac{\omega}{F}}\,,
\label{hp}
\end{eqnarray}
the action (\ref{Einstein}) can be written in the canonical form
\begin{eqnarray}
{\cal S}_E=\int d^{4} \hat{x} \sqrt{-\hat{g}} 
\left[ \frac12 \hat{R}-\frac12 
(\nabla \hp)^2-\hV(\phi) \right]\,.
\label{Eincan}
\end{eqnarray}

We shall consider a perturbed metric in the Einstein frame
\begin{eqnarray}
d\hat{s}^2 &=& \Omega ds^2 \\ 
&=&-(1+2\hat{A})d\hat{t}^2 +2\hat{a}(\hat{t})
\hat{B}_{,i}d\hat{x}^i d\hat{t}
+\hat{a}^2(\hat{t})\left[(1+2\hat{\varphi})\delta_{ij} 
+2\hat{E}_{,i,j}\right]d\hat{x}^i d \hat{x}^j\,,
\label{pmetricEin}
\end{eqnarray}
and decompose the conformal factor into the background and 
the perturbed part as
\begin{eqnarray}
\Omega({\bf x}, t)=\bar{\Omega}(t) 
\left(1+\frac{\delta \Omega({\bf x}, t)}
{\bar{\Omega}(t)}\right)\,.
\label{Omega}
\end{eqnarray}
In what follows we drop a ``bar'' when we express $\bar{\Omega}(t)$.
Then we get the following relations 
\begin{eqnarray}
\hat{a}=a\sqrt{\Omega},~~d\hat{t}=\sqrt{\Omega}dt,~~
\hat{H}=\frac{1}{\sqrt{\Omega}}
\left(H+\frac{\dot{\Omega}}{2\Omega}\right),~~
\hat{A}=A+\frac{\delta \Omega}{2\Omega},~~
\hat{\varphi}=\varphi+\frac{\delta \Omega}{2\Omega}.
\label{relation}
\end{eqnarray}

Making use of these relations, it is easy to show that 
curvature perturbations in the Einstein frame exactly 
coincide with those in the Jordan frame:
\begin{eqnarray}
\hat{\cal R} &\equiv& \hat{\varphi}-\frac{\hat{H}}
{{\rm d}\hp/{\rm d}\hat{t}}\delta \hat{\phi} \\
&=& \varphi-\frac{H}{\dot{\phi}}\delta\phi={\cal R}\,.
\label{calRein}
\end{eqnarray}
Since tensor perturbations are also invariant 
under a conformal transformation, we have the following relations
\begin{eqnarray}
\hat{\cal P}_{\rm S}={\cal P}_{\rm S}\,,~~~~
\hat{\cal P}_{\rm T}={\cal P}_{\rm T}\,.
\label{RT}
\end{eqnarray}

Introducing the following quantities
\begin{eqnarray}
\hat{\gamma}_{\rm S}=
\frac{(1+\hat{\delta}_{\rm S})
(2+\hat{\delta}_{\rm S}+\hat{\epsilon})}
{(1+\hat{\epsilon})^2}\,,~~~
\hat{\epsilon}=\frac{{\rm d}\hat{H}/{\rm d}\hat{t}}{\hat{H}^2}\,,
~~~\hat{\delta}_{\rm S}=\frac{{\rm d}\hat{Q}_{\rm S}/{\rm d}\hat{t}}
{2\hat{H}\hat{Q}_{\rm S}}\,,~~~
\hat{Q}_{\rm S}=\left(\frac{{\rm d}\hp/{\rm d}\hat{t}}
{\hat{H}}\right)^2=Q_{\rm S}/F\,,
\end{eqnarray}
the spectral index of scalar perturbations in the 
Einstein frame is given by 
\begin{eqnarray}
\label{tilns}
\hat{n}_{\rm S}-1=3-\sqrt{4\hat{\gamma}_{\rm S}+1}\,.
\end{eqnarray}
The two quantities $\hat{\epsilon}$ and  $\hat{\delta}_{\rm S}$
can be written as 
\begin{eqnarray}
\hat{\epsilon}=\frac{\epsilon-\beta}{1+\beta}+
\frac{\dot{\beta}}{H(1+\beta)^2}\,,~~~
\hat{\delta}_{\rm S}=\frac{\delta-\beta}{1+\beta}\,,~~~
{\rm with}~~~\beta=\frac{\dot{F}}{2HF}\,.
\label{apdel}
\end{eqnarray}
When the variation of $\beta$ is negligible 
($\dot{\beta} \simeq 0$), which is valid in the context of slow-roll 
inflation, one can easily show that 
$\hat{\gamma}_{\rm S}=\gamma_{\rm S}$.
Therefore the spectral index (\ref{tilns}) in the Einstein frame 
coincides with the one in the Jordan frame 
($\hat{n}_{\rm S}=n_{\rm S}$).
The spectral index of tensor perturbations is also the same in both frames
($\hat{n}_{\rm T}=n_{\rm T}$), and simply given as 
\begin{eqnarray}
\hat{n}_{\rm T}=3-\sqrt{4\hat{\gamma}_{\rm T}+1}\,,
~~~{\rm with}~~~\hat{\gamma}_{\rm T}=
\frac{2+\hat{\epsilon}}{(1+\hat{\epsilon})^2}\,,
\label{nTein}
\end{eqnarray}
where we used $\hat{Q}_{\rm T}=Q_{\rm T}/F=1$ 
and $\hat{\delta}_{\rm T}=0$.

The tensor-to-scalar ratio is unchanged 
by a conformal transformation and is given by 
\begin{eqnarray}
\hat{R}=\frac{\hat{A}_{\rm T}^2}{\hat{A}_{\rm S}^2}
= 8\frac{\hat{Q}_{\rm S}}{\hat{Q}_{\rm T}}
\left(\frac{\Gamma(\hat{\nu}_{\rm T})}
{\Gamma(\hat{\nu}_{\rm S})} \right)^2
=8\frac{Q_{\rm S}}{Q_{\rm T}}
\left(\frac{\Gamma(\nu_{\rm T})}
{\Gamma(\nu_{\rm S})} \right)^2=R\,,
\label{ratioEin}
\end{eqnarray}
where we used $\hat{Q}_{\rm S}=Q_{\rm S}/F$,
$\hat{Q}_{\rm T}=Q_{\rm T}/F$, 
$\hat{\nu}_{\rm S}=\nu_{\rm S}$ and 
$\hat{\nu}_{\rm T}=\nu_{\rm T}$.
Making use of the background equation, 
$2{\rm d}\hat{H}/{\rm d}\hat{t}
=-({\rm d}\hp/{\rm d}\hat{t})^2$, we get
\begin{eqnarray}
\hat{R}=8\left(\frac{{\rm d}\hp/{\rm d}\hat{t}}
{\hat{H}}\right)^2
\left(\frac{\Gamma(\hat{\nu}_{\rm T})}
{\Gamma(\hat{\nu}_{\rm S})} \right)^2
=-16 \frac{{\rm d}\hat{H}/{\rm d}\hat{t}}{\hat{H}^2}
\left(\frac{\Gamma(\hat{\nu}_{\rm T})}
{\Gamma(\hat{\nu}_{\rm S})} \right)^2\,.
\label{ratioep}
\end{eqnarray}
Hereafter we shall drop a ``hat'' for the quantities $\hat{n}_{\rm S}$,
$\hat{n}_{\rm T}$, $\hat{R}$.

\subsection{Slow-roll analysis}

If the slow-roll approximation is employed 
($|\epsilon| \ll 1$, $|\delta_{\rm S}| \ll1$, 
$|\delta_{\rm T}| \ll1$ and $|\beta| \ll 1$), 
we get the following relations
\begin{eqnarray}
\label{slowns}
n_{\rm S} &=& 1 +2\epsilon-2\delta_{\rm S}=
1+2\hat{\epsilon}-2\hat{\delta}_{\rm S}\,, \\
n_{\rm T} &=& 2\epsilon-2\delta_{\rm T}=
2\hat{\epsilon}\,, \\
R &=& -8n_{\rm T}\,.
\label{slow}
\end{eqnarray}
Here we used $\Gamma(\nu_{\rm S}) \simeq \Gamma(\nu_{\rm T})
\simeq \Gamma(3/2)$, 
since the spectra of scalar and tensor perturbations are 
close to scale-invariant. The consistency relation (\ref{slow}) is 
the same as in the case of the standard
Einstein gravity\footnote{Note that the consistency relation is 
different in the context of multi-field inflation 
due to the correlation between adiabatic and isocurvature
perturbations. See Refs.\,\cite{cor} for details.}. 
This means that a separate likelihood analysis of 
observational data is not needed even for the generalized action
(\ref{lag}), when we vary four observational quantities
$n_{\rm S}$, $n_{\rm T}$, $R$ and $A_{\rm S}^2$.
This situation is similar to the perturbations in 
Randall-Sundrum II braneworld scenario 
in which the same consistency 
relation ({\ref{slow}) holds \cite{Huey}.
Thus we can exploit the observational constraints on the values of 
$n_{\rm S}$ and $R$ which were recently derived in 
Ref.~\cite{TL}.

On the other hand, when we constrain the inflaion models which 
belong to the generalized action (\ref{lag}), the difference is seen 
compared to the Einstein gravity.
This comes from the fact that the potential in the Einstein 
frame includes the conformal factor $1/F^2$ and that the 
field $\hat{\phi}$ is different from the inflaton $\phi$
[see Eq.~(\ref{hp})].
Under the slow-roll approximation, the background equations 
in the Einstein frame can be written as
\begin{eqnarray}
\hat{H}^2 \simeq \kappa^2\hat{V}\,,~~~
3\hat{H}\frac{{\rm d}\hat{\phi}}{{\rm d}\hat{t}}
\simeq -\hat{V}_{\hp}\,.
\label{backslow}
\end{eqnarray}
Then $\hat{\epsilon}$ and $\hat{\delta}_{\rm S}$ are written 
in terms of the slope of the potential
\begin{eqnarray}
\hat{\epsilon} \simeq -\frac{1}{2\kappa^2}
\left(\frac{\hV_{\hp}}{\hV}\right)^2\,,~~~~
\hat{\delta}_{\rm S} \simeq 
\frac{1}{\kappa^2}
\left[\left(\frac{\hV_{\hp}
}{\hV}\right)^2-
\frac{\hV_{\hp \hp}}{\hV}\right]\,.
\label{slope}
\end{eqnarray}
Then we can evaluate the theoretical values $n_{\rm S}$, $n_{\rm T}$, 
$R$ and constrain model parameters by 
comparing with observational data.

\section{The nonminimally coupled inflaton field} 

In this section we wish to apply the formalism in the previous 
section to a nonminimally coupled inflaton field.
In this case we have 
\begin{eqnarray}
F(\phi)=(1-\xi \kappa^2\phi^2)/\kappa^2\,,~~~
\omega(\phi)=1\,,
\label{nonmin}
\end{eqnarray}
for the action (\ref{lag}).
Here we used the notation where the conformal coupling 
corresponds to $\xi=1/6$, which is the same notation 
as Futamase and Maeda \cite{FM}
but different from Fakir and Unruh \cite{FU}.

\subsection{Background equations}

In the case (\ref{nonmin}) the background equations 
(\ref{back2})-(\ref{back}) are written in the form 
\begin{eqnarray}
& &H^2=\frac{\kappa^2}{6(1-\xi \kappa^2 \phi^2)}
\left[\dot{\phi}^2+2V+12H\xi \phi \dot{\phi}\right]\,, \\
& & \ddot{\phi}+3H\dot{\phi}-\frac{\xi (1-6\xi) \kappa^2 \phi \dot{\phi}^2}
{1-(1-6\xi)\xi \kappa^2 \phi^2}+
\frac{4\xi\kappa^2 \phi V+(1-\xi \kappa^2 \phi^2)V_\phi}
{1-(1-6\xi)\xi  \kappa^2 \phi^2}=0\,,
\label{basiceq}
\end{eqnarray}
where we used $R=6(2H^2+\dot{H})$.

Hereafter we shall consider the large-field potential (\ref{potential}).
Under the slow-roll approximations, $|\ddot{\phi}| \ll |3H\dot{\phi}|$
and $|3H\dot{\phi}| \ll |V_\phi|$, one has
\begin{eqnarray}
\label{slowroll2}
& & 3H\dot{\phi} \simeq -\frac{c\phi^{p-1}
[p+\psi(4-p)]}{1-(1-6\xi)\psi}\,, \\
& & H^2 \simeq \frac{\kappa^2 c\phi^p}{3(1-\psi)}
\left[1-\frac{2\xi \{p+\psi(4-p)\}}
{1-(1-6\xi)\psi}\right]\,.
\label{slowroll}
\end{eqnarray}
where 
\begin{eqnarray}
\psi \equiv \xi \kappa^2 \phi^2\,.
\label{ps}
\end{eqnarray}
We get the following equation from 
Eqs.~(\ref{slowroll2}) and (\ref{slowroll})
\begin{eqnarray}
\frac{\dot{\psi}}{H}=
-\frac{2\xi(\psi-1)\left[(p-4)\psi-p \right]}
{(2\xi p-2\xi -1)\psi+1-2\xi p}\,.
\label{dotpsi}
\end{eqnarray}
The number of $e$-folds is defined as
\begin{eqnarray}
N \equiv \int_t^{t_{\rm f}} H {\rm d}t
=\int_{\psi}^{\psi_{\rm f}}
\frac{H}{\dot{\psi}}{\rm d}\psi\,,
\label{efolddef}
\end{eqnarray}
where the subscript ``${\rm f}$'' denotes the values at the end of
inflation. Making use of Eq.~(\ref{dotpsi}), we have
\begin{eqnarray}
\label{efold2}
N &=& -\frac{1}{4\xi}
{\rm ln}\, \left|\frac{(p-4)\psi_{\rm f}-p}{(p-4)\psi-p}
\right|^{\frac{3\xi p-2}{p-4}}
\left|\frac{\psi_{\rm f}-1}{\psi-1}\right|^{\xi}\,,~~
{\rm for}~~p \ne 4\,, \\
N &=& -\frac{1-6\xi}{8\xi} \left(\psi_{\rm f}-\psi \right)
-\frac14 {\rm ln}\,\left|\frac{\psi_{\rm f}-1}{\psi-1}\right|\,,
~~{\rm for}~~p=4\,.
\label{efold}
\end{eqnarray}
Later we shall use this relation to express $\psi$ in terms
of $N$.

\subsection{The potential in the Einstein frame}

The potential in the Einstein frame and the effective gravitational constant 
are given, respectively, by
\begin{eqnarray}
\hV=\frac{c\phi^p}{(1-\xi \kappa^2\phi^2)^2}\,,~~~~
G_{\rm eff}=\frac{G}{1-\xi \kappa^2 \phi^2}\,.
\label{hatv}
\end{eqnarray}
When $\xi$ is positive, we need the condition $\phi^2<\phi_c^2 \equiv
m_{\rm pl}^2/(8\pi \xi)$ in order to reproduce the present value of
the gravitational constant. Futamase and Maeda obtained the 
constraint $\xi \lesssim 10^{-3}$ from the requirement that the 
initial value of the inflaton ($\phi_i \sim 5m_{\rm pl}$)
is smaller than $\phi_c$ \cite{FM}.
They also pointed out that this constraint becomes more stringent 
when $\phi_i$ is larger.

We do not have the singularities of the potential $\hV$
and the effective gravitational constant 
for negative values of $\xi$.
Futamase and Maeda found that the negative
values of $\xi$ are generically allowed for $p=4$,
while the nonminimal coupling is constrained to be 
$|\xi| \lesssim 10^{-3}$ for $p=2$ from the requirement of 
a sufficient inflation. 
In particular Fakir and Unruh \cite{FU} showed that the fine tuning of 
the coupling constant $c$ is avoided for $p=4$ by considering large 
negative values of $\xi$ satisfying $|\xi| \gg 1$.
Hereafter we shall concentrate on the case of negative $\xi$
for the general potential (\ref{potential}).
We wish to constrain the strength of nonminimal couplings 
by exploiting latest observational data.
While Komatsu and Futamase \cite{KF0,KF} 
focused on large negative nonminimal 
couplings for $p=4$, we shall consider general 
values of $\xi$ with the generic potential (\ref{potential}).

In Fig.\,\ref{po} we plot the potential $\hV$ in the Einstein frame 
for $p=2$, $p=4$ and $p=6$ when $\xi$ is negative.
{}From Eq.~(\ref{hatv}) the potential has  a local maximum at 
\begin{eqnarray}
\phi_{\rm M}=\sqrt{\frac{p}{8\pi (4-p)|\xi|}}\,m_{\rm pl}\,,
\label{phiM}
\end{eqnarray}
as long as $p<4$. 
Therefore we have $\phi_{\rm M}=m_{\rm pl}/\sqrt{8\pi |\xi|}$
for $p=2$. Naively we expect that $\phi_{\rm M}$ is required 
to be larger than $3m_{\rm pl}$ in order to lead to a sufficient number of 
$e$-folds ($N \gtrsim 60$), which gives $|\xi| \lesssim 4.4 \times 10^{-3}$.
Note, however, that one can get a large number of $e$-folds
if the initial value of $\phi$ is close to $\phi_{\rm M}$.
Therefore nonminimal couplings satisfying $|\xi| \gtrsim 4.4 \times 10^{-3}$
may not be excluded\footnote{In Ref.~\cite{SY} it was shown that the 
constraint on $\xi$ is relaxed by considering topological inflation.}.
In this work we will place complete constraints on the strength of $\xi$
by comparing the perturbation spectra with observations.

When $p=4$ the potential $\hV$ monotonically increases toward a 
constant value, $\hV \to c/\xi^2\kappa^4$, as $\phi \to \infty$.
Since the potential becomes flatter by taking into account negative 
nonminimal couplings, the amount of inflation gets larger compared 
to the case of $\xi=0$.
For $|\psi| \gg 1$, we obtain the following background 
solution from Eqs.~(\ref{slowroll2}) and (\ref{slowroll}), as
\begin{eqnarray}
\phi=\phi_0-\frac{4\sqrt{c}}{(1-6\xi)\sqrt{-3\xi}\kappa^2}t\,,~~~~
a=a_0 \exp\left[\sqrt{\frac{c}{-3\xi}}
\left(\phi_0 t-\frac{2\sqrt{c}}{(1-6\xi)\sqrt{-3\xi}\kappa^2}t^2
\right)\right]\,,
\label{p=4back}
\end{eqnarray}
where $\phi_0$ and $a_0$ are constants.
This means that the universe expands quasi-exponentially 
for large negative values of $\xi$.
 
In the case of $p>4$, the steepness of the potential 
$\hat{V}$ is relaxed by taking into account negative $\xi$.
If $|\psi|$ is much smaller than unity, 
the effect of nonminimal couplings helps to lead to 
a larger number of $e$-folds.
When $|\psi| \gg 1$, we have the following analytic solution 
\begin{eqnarray}
\phi=\left[\frac{(p-2)(p-4)\sqrt{c}}{12\sqrt{-\xi(p-1)}}t
+\phi_0 \right]^{\frac{2}{2-p}}\,,~~~~
a=a_0 \left[t-\frac{6\xi \phi_0}{c(4-p)}\right]
^{\frac{4(p-1)}{(p-2)(p-4)}}\,.
\label{pnon4back}
\end{eqnarray}
This explicitly shows that 
the field $\phi$ decreases with time for $p>4$.
We also find that the solution (\ref{pnon4back}) does not correspond
to an inflationary solution for $p>5+\sqrt{13}$.

The slow-roll parameter in the Einstein frame is given as
\begin{eqnarray}
|\hat{\epsilon}|=\frac{1}{2\kappa^2 G^2(\phi)}
\left(\frac{\hV_{\phi}}{\hV}\right)^2
=\frac{\xi}{2\psi}
\frac{[p+(4-p)\psi]^2}
{1-(1-6\xi)\psi}\,,
\label{slownonmini}
\end{eqnarray}
where we used $G^2(\phi)=[1-(1-6\xi)\psi]/(1-\psi)^2$.
Therefore one has $|\hat{\epsilon}|=(p-4)^2/12$ for $|\psi| \gg 1$
and $|\xi| \gg 1$, 
which means that $|\hat{\epsilon}|$ is larger than unity for
$p>4+2\sqrt{3}$.
This again shows that inflationary solutions are not obtained
for $p\gtrsim 8$. Since we can not keep the 
$1/G^2(\phi)$ term to be small in Eq.~(\ref{slownonmini})
for $|\psi| \gg 1$, the slow-roll parameter exceeds unity unless $p$
is small. 
Hereafter we shall mainly investigate the cases of $p=2$, $p=4$
and $p=6$. This is sufficient to understand what happens for 
the perturbation spectra in the presence of
nonminimal couplings.

\begin{figure}
\begin{center}
\singlefig{10cm}{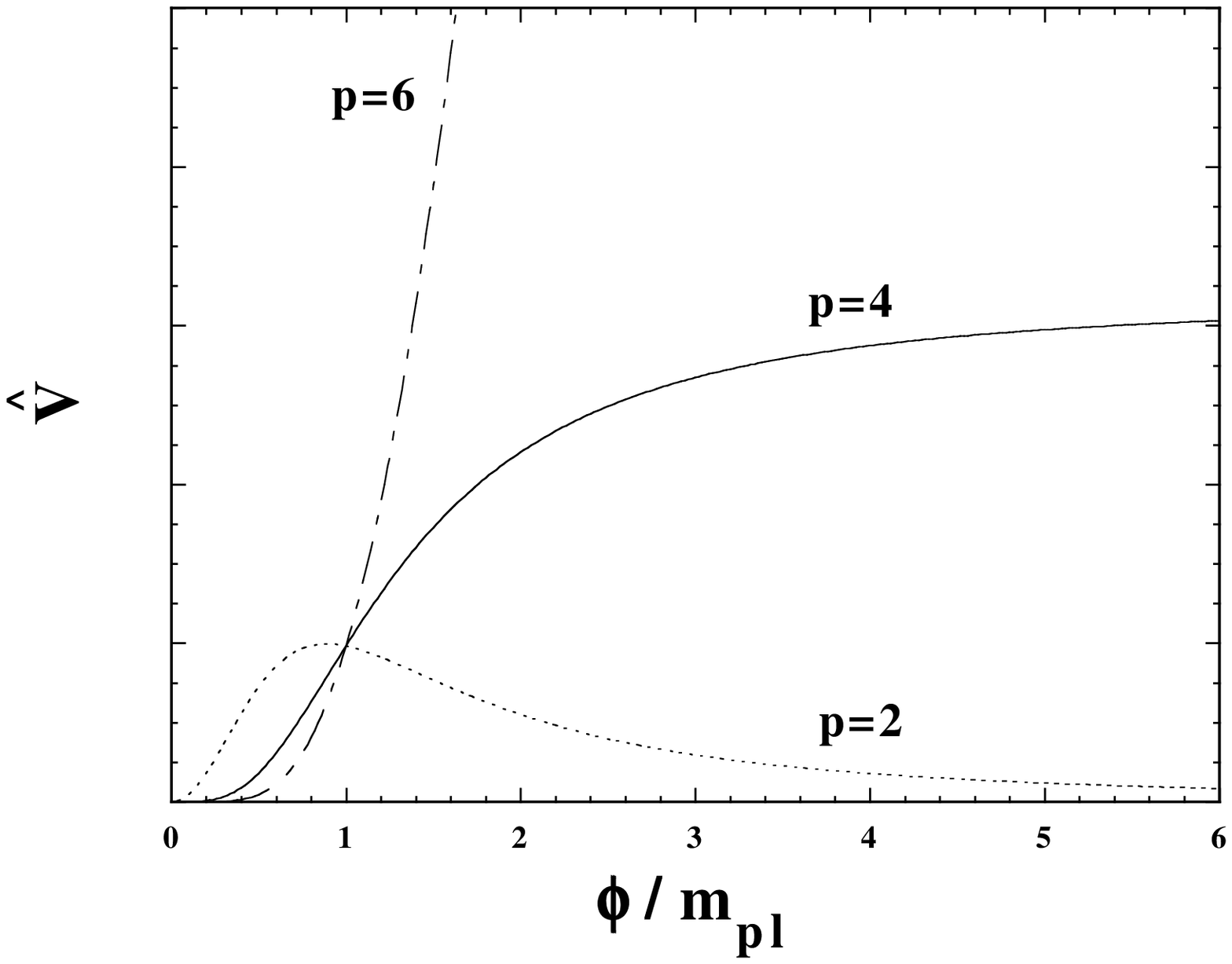}
\begin{figcaption}{po}{10cm}
The potential of the inflaton in the Einstein frame 
with $\xi=-0.05$ for $p=2$, $p=4$ and $p=6$.
It has a local maximum 
at $\phi_{\rm M}=m_{\rm pl}/\sqrt{8\pi |\xi|}$
for $p=2$.
When $p=4$ the potential approaches
a constant value $\hV=c/\xi^2 \kappa^4$ 
as $\phi \to \infty$.
\end{figcaption}
\end{center}
\end{figure}

\subsection{Perturbation spectra and the tensor-to-scalar ratio}

We are now in the stage to evaluate the spectra of perturbations
for the nonminimally coupled inflaton field.
Making use of the results (\ref{slowns}) and (\ref{slow})
with slow-roll parameters (\ref{slope}), 
we get the values of $n_{\rm S}$ and $R$ as
\begin{eqnarray}
\label{nSandRd}
n_{\rm S}-1 &=&\frac{\xi}{\psi}
\frac{p+(4-p)\psi}{1-(1-6\xi)\psi} \nonumber \\
& & \times \left[-3p+(3p-8)\psi+\frac{2(1-6\xi)\psi(1-\psi)}
{1-(1-6\xi)\psi}+2\frac{(1-\psi)^2p(p-1)+4\psi \left\{
1+p+(1-p)\psi \right\}}{p+(4-p)\psi}\right]\,, \\
R &=& \frac{8\xi}{\psi}\frac{[p+(4-p)\psi]^2}
{1-(1-6\xi)\psi}\,,
\label{nSandR}
\end{eqnarray}
which are written in terms of the functions of $\xi$ and $\psi$
for a fixed value of $p$.

The end of inflation is characterized by $|\hat{\epsilon}|=1$,
thereby yielding 
\begin{eqnarray}
\psi_{\rm f}=\frac{1-\xi p(4-p)-\sqrt{(1-2p\xi)(1-6p\xi)}}
{\xi (4-p)^2+2(1-6\xi)}\,,
\label{psif}
\end{eqnarray}
which we choose the negative sign of $\psi_{\rm f}$, since 
we are considering the case of $\xi<0$.
When $|\xi| \ll 1$, Eqs.\,(\ref{efold2}) and 
(\ref{efold}) give the following relation 
\begin{eqnarray}
\label{psiapp2}
& &\left|(p-4)\psi-p \right| \simeq \left|(p-4)
\psi_{\rm f}-p \right|
\exp\left[-2\xi N(p-4) \right]\,,~~~
{\rm for}~~~p \neq 4\,, \\
& & \psi \simeq \psi_{\rm f}+
\frac{8\xi}{1-6\xi}N\,,~~~
{\rm for}~~~p=4\,.
\label{psiapp}
\end{eqnarray}
Making use of Eqs.~(\ref{psiapp2}) and (\ref{psiapp}),
one can express $n_{\rm S}$ and $R$
in terms of $\xi$ and $N$.
Fixing the $e$-fold at the cosmologically relevant scale
$N=55$, $n_{\rm S}$ and $R$ are the function of 
$\xi$ only. Therefore we can constrain the strength of 
nonminimal couplings by comparing the theoretical predictions
(\ref{nSandRd}) and (\ref{nSandR}) with observational data.

Before proceeding the constraint on $\xi$, we shall investigate
the effect of nonminimal couplings under the approximations of
$|\xi| \ll 1$ and $|\psi| \ll 1$. In this case we have 
\begin{eqnarray}
\label{nRad}
n_{\rm S}-1 & \simeq & -\frac{\xi p(p+2)}{\psi}
\left[1-\frac{p^2-8p+8}{p(p+2)}\psi \right]\,, \\
R & \simeq & \frac{8p^2\xi}{\psi}
\left(1+\frac{8-p}{p}\psi\right)\,.
\label{nRap}
\end{eqnarray}
In what follows we shall investigate the cases of $p=2$, $p=4$
and $p=6$ separately.

\subsubsection{Case of $p=2$}

In this case one has the following relation from 
Eqs.~(\ref{psiapp2}) and (\ref{psif}):
\begin{eqnarray}
\psi=\left|\frac32-\frac12 \sqrt{\frac{1-12\xi}{1-4\xi}}
\right| e^{4\xi N}-1 \simeq e^{4\xi N}-1\,,
\label{psip=2}
\end{eqnarray}
which is valid for $|\xi| \ll 1$.
Then Eqs.~(\ref{nRad}) and (\ref{nRap}) reduce to 
\begin{eqnarray}
n_{\rm S}-1 &\simeq& -4\xi \frac{e^{4\xi N}+1}
{e^{4\xi N}-1} \simeq -\frac{2}{N}
\left(1+\frac43 \xi^2N^2\right)\,, \\
R &\simeq& 32\xi \frac{3e^{4\xi N}-2}
{e^{4\xi N}-1} \simeq \frac{8}{N}
\left(1+10\xi N\right)\,.
\label{nRp=2}
\end{eqnarray}
In the minimally coupled case ($\xi=0$), we have 
$n_{\rm S}=1-2/N$ and $R=8/N$, corresponding to 
$n_{\rm S}=0.964$ and $R=0.145$ for $N=55$.
If we take into account negative nonminimal couplings, 
we find the increase of $|n_{\rm S}-1|$ and the decrease of 
$R$ compared to the case of $\xi=0$.
This behaviour is clearly seen in Figs.~\ref{ns} and 
\ref{R} that are obtained without using the approximation above.
The rapid decrease of $R$ for $|\xi| \gtrsim 10^{-3}$ reflects the 
fact that $\phi$ approaches the value $\phi_{\rm M}$
for larger $|\xi|$, which works to decrease
$|\hat{\epsilon}|$ toward zero. 
On the other hand, the second derivative of the potential
gets larger on the R.H.S. of $\hat{\delta}_{\rm S}$
with the growth of $|\xi|$.
This is the reason why $n_{\rm S}$ departs from 1
for larger $|\xi|$ in spite of the fact that $\hat{\epsilon}$
gets smaller toward zero. 

\begin{figure}
\begin{center}
\singlefig{10cm}{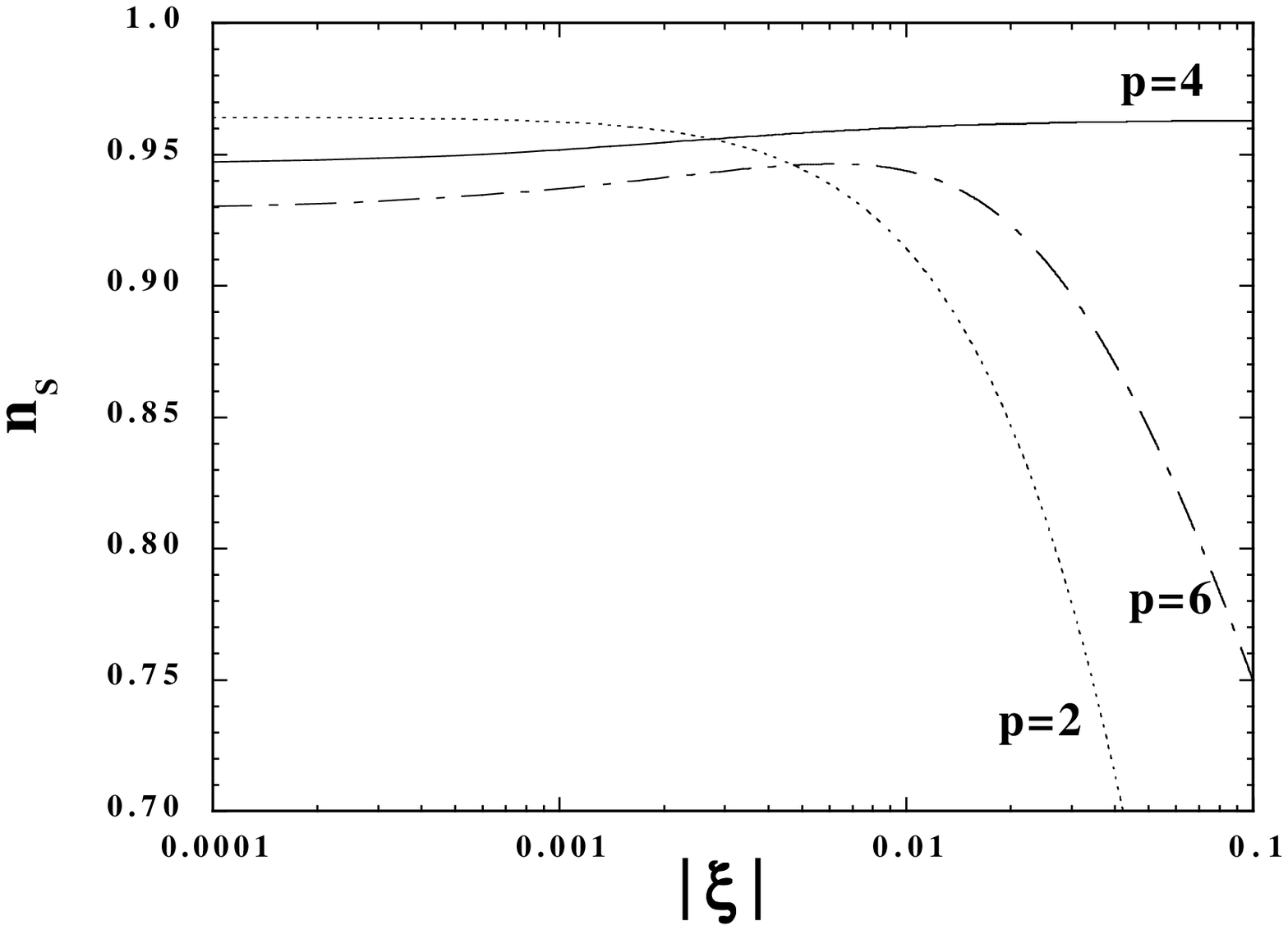}
\begin{figcaption}{ns}{10cm}
The spectral index $n_{\rm S}$ as a function of $|\xi|$
for $p=2$, $p=4$ and $p=6$. Note that we are considering 
negative values of $\xi$. See the text for the interpretation 
of this figure.
\end{figcaption}
\end{center}
\end{figure}

\begin{figure}
\begin{center}
\singlefig{10cm}{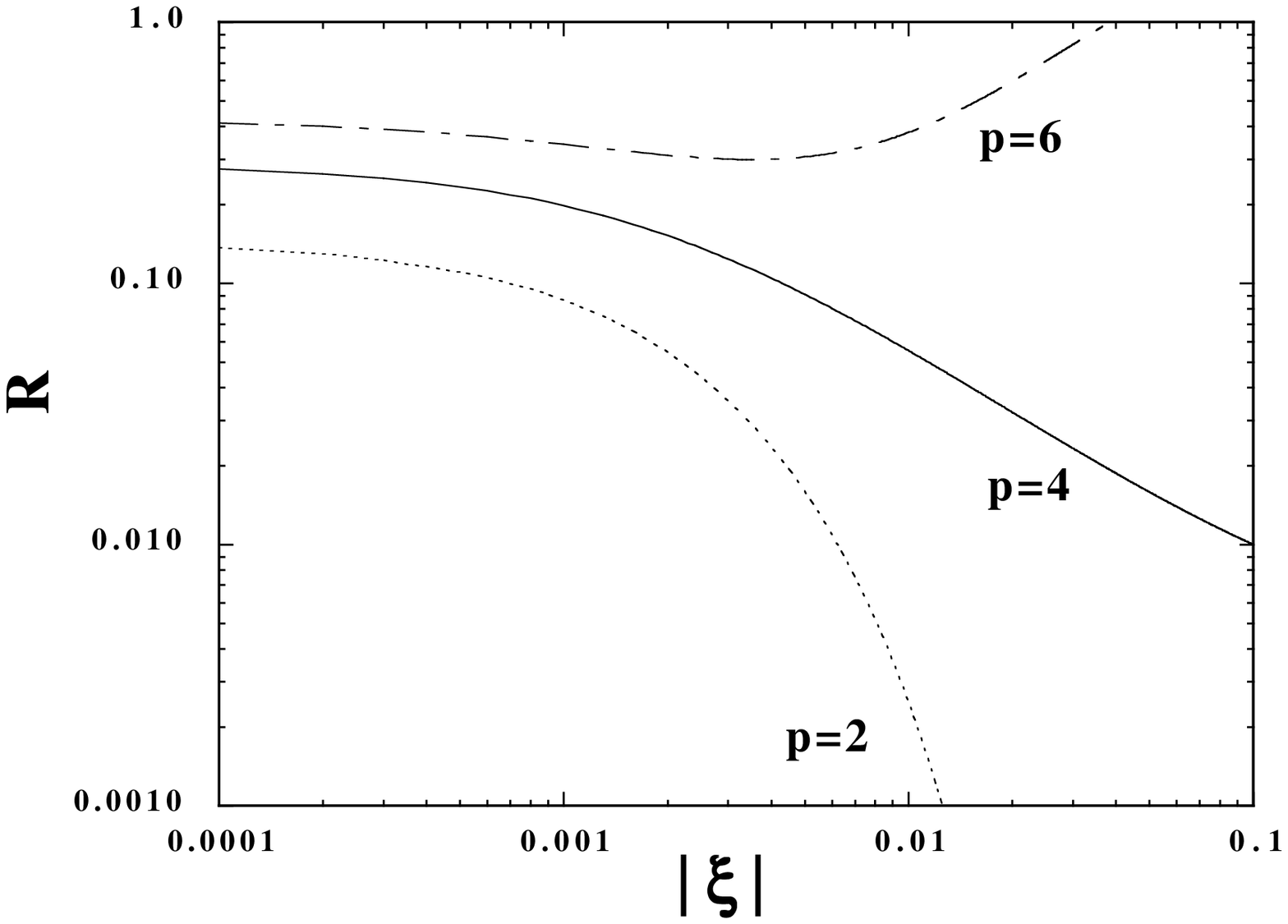}
\begin{figcaption}{R}{10cm}
The tensor-to-scalar ratio $R$ as a function of $|\xi|$
for $p=2$, $p=4$ and $p=6$ with negative $\xi$.
\end{figcaption}
\end{center}
\end{figure}

\subsubsection{Case of $p=4$}

In the case of the quartic potential, 
Eqs.~(\ref{psiapp}) and (\ref{psiapp})
give the following relation
\begin{eqnarray}
\psi=\frac{1-\sqrt{(1-8\xi)(1-24\xi)}+16\xi N}{2(1-6\xi)}\,.
\label{psip=4}
\end{eqnarray}
Note that this is valid for general values of $\xi$, since the 
second term on the R.H.S. of Eq.~(\ref{efold}) is always 
sub-dominant relative to the first term.
When the condition, $|\xi| \ll 1$, is satisfied, one has 
$\psi \simeq 8\xi N$ from Eq.~(\ref{psip=4}).
If $|\psi|$ is smaller than of order unity, we find 
\begin{eqnarray}
n_{\rm S}-1 &\simeq& -\frac{24\xi}{\psi}
\left(1+\frac13 \psi \right) \simeq
-\frac{3}{N} \left(1+\frac83 \xi N \right)\,, \\
R &\simeq& \frac{128\xi}{\psi}
\left(1+\psi\right) \simeq \frac{16}{N}
\left(1+8 \xi N \right) \,.
\label{nRp=4}
\end{eqnarray}
Then we have $n_{\rm S}=1-3/N$ and $R=16/N$ for 
$\xi \to 0$, corresponding to $n_{\rm S}=0.945$
and $R=0.291$ for $N=55$.
Inclusion of nonminimal couplings leads to 
the decrease of both $|n_{\rm S}-1|$ and $R$, as seen 
in Figs.\,\ref{ns} and \ref{R}.
This comes from the fact that the potential becomes flatter
in the presence of negative nonminimal couplings.

It is worth mentioning the case of large negative nonminimal 
couplings ($|\xi| \gg 1$).
Since $\hat{\epsilon} \simeq 8\xi/(1-6\xi)\psi^2$ and 
$\hV_{\hat{\phi}\hat{\phi}}/\kappa^2 \hV \simeq 
-8\xi/(1-6\xi)\psi$ in this case, we find 
\begin{eqnarray}
n_{\rm S}-1 \simeq -\frac{16\xi}{1-6\xi}
\frac{1}{\psi}\,,~~~~
R \simeq \frac{-128\xi}{1-6\xi}
\frac{1}{\psi^2}\,.
\label{nRp=4d}
\end{eqnarray}
{}From Eq.\,(\ref{psip=4}) one obtains $\psi \simeq -4N/3$
for $|\xi| \gg 1$. Then we get the following results
\begin{eqnarray}
n_{\rm S}-1 \simeq -2/N\,,~~~R \simeq 12/N^2\,.
\label{nRp=4limit}
\end{eqnarray}
Note that the spectral index $n_{\rm S}$ is the same as
in the minimally coupled 
case with the quadratic potential ($\xi=0$ and $p=2$).
One has $n_{\rm S}=0.964$ and $R=0.00397$ for $N=55$
\footnote{Komatsu and Futamase \cite{KF} obtained the values
$n_{\rm S}=0.97$ and $R=0.002$, since they choosed the 
$e$-fold $N=70$.}.
Therefore the Fakir and Unruh scenario
with $|\xi| \gg 1$ predicts a much smaller 
value of $R$ compared to the minimally coupled case.
 
\subsubsection{Case of $p=6$}

For $p=6$ and $|\xi| \ll 1$, one obtains the following relation 
from Eq.~(\ref{psiapp2}) 
\begin{eqnarray}
\psi=3-|3-\psi_{\rm f}|e^{-4\xi N}\,,
\label{psip=6}
\end{eqnarray}
where $\psi_{\rm f}$ is approximately given as 
$\psi_{\rm f} \simeq 18\xi$ by Eq.~(\ref{psif}).
Then we find $\psi \simeq 12\xi N$ and 
\begin{eqnarray}
n_{\rm S}-1 &\simeq& -\frac{48\xi}{\psi}
\left(1+\frac{1}{12}\psi \right) \simeq 
-\frac{4}{N}(1+\xi N)\,, \\
R &\simeq& \frac{288\xi}{\psi}
\left(1+\frac{1}{3}\psi \right)\simeq 
-\frac{24}{N}(1+4\xi N)\,.
\label{nRp=6}
\end{eqnarray}
This indicates that negative nonminimal couplings lead to 
the decrease of $|n_{\rm S}-1|$ and $R$ for $|\xi| \ll 1$.

From Figs.\,\ref{ns} and \ref{R} we find that 
$|n_{\rm S}-1|$ and $R$ begin to increase for 
$|\xi| \gtrsim 10^{-2}$.
This can be understood as follows.
When $p \ne 4$ and $|\psi|$ is larger than of order unity, 
we obtain
\begin{eqnarray}
n_{\rm S}-1 &\simeq& 
\frac{4\xi}{1-6\xi}(p-4)^2\,, \\
R &\simeq& 
\frac{-16\xi}{1-6\xi}(p-4)^2\,,
\label{nRps}
\end{eqnarray}
which are independent of $N$.
In the case of $p=6$, this yields $n_{\rm S}-1=16\xi/(1-6\xi)$
and $R=-64\xi/(1-6\xi)$. Therefore both $|n_{\rm S}-1|$ and 
$R$ grow with the increase of $|\xi|$.
The asymptotic values correspond to $n_{\rm S}-1 \to -8/3$ 
and $R \to 32/3$ as $|\xi| \to \infty$.

\subsection{Observational constraints on nonminimal couplings}

Lets us now place observational constraints on the strength of 
nonminimal couplings.
As shown in Sec.\,II, the inflationary observables $P_{\rm S}$,
$R$, $n_{\rm S}$ and $n_{\rm T}$ are equivalent both in 
the Jordan frame and the Einstein frame.
This correspondence indicates that a separate likelihood analysis of 
observational data is not required compared to the Einstein gravity.
Recently one of the present authors carried out a  likelihood analysis \cite{TL}
in the context of braneworld inflation using a compilation of 
data including 
WMAP \cite{WMAP1,WMAP2,WMAP3}, the 2dF \cite{2dF} and 
latest SDSS galaxy redshift surveys \cite{SDSS}.
Since the same correspondence holds for inflationary observables in this 
case as well, we can exploit the observational constraints derived in 
Ref.\,\cite{TL}. Note that we used the 
CosmoMc (Cosmological Monte Carlo) code \cite{antony2}
with the CAMB program \cite{antony1}, and varied four inflationary
variables in addition to four cosmological parameters 
($\Omega_b h^2$, $\Omega_c h^2$, $Z=e^{-2\tau}$, $H_0$).

\begin{figure}
\begin{center}
\singlefig{10cm}{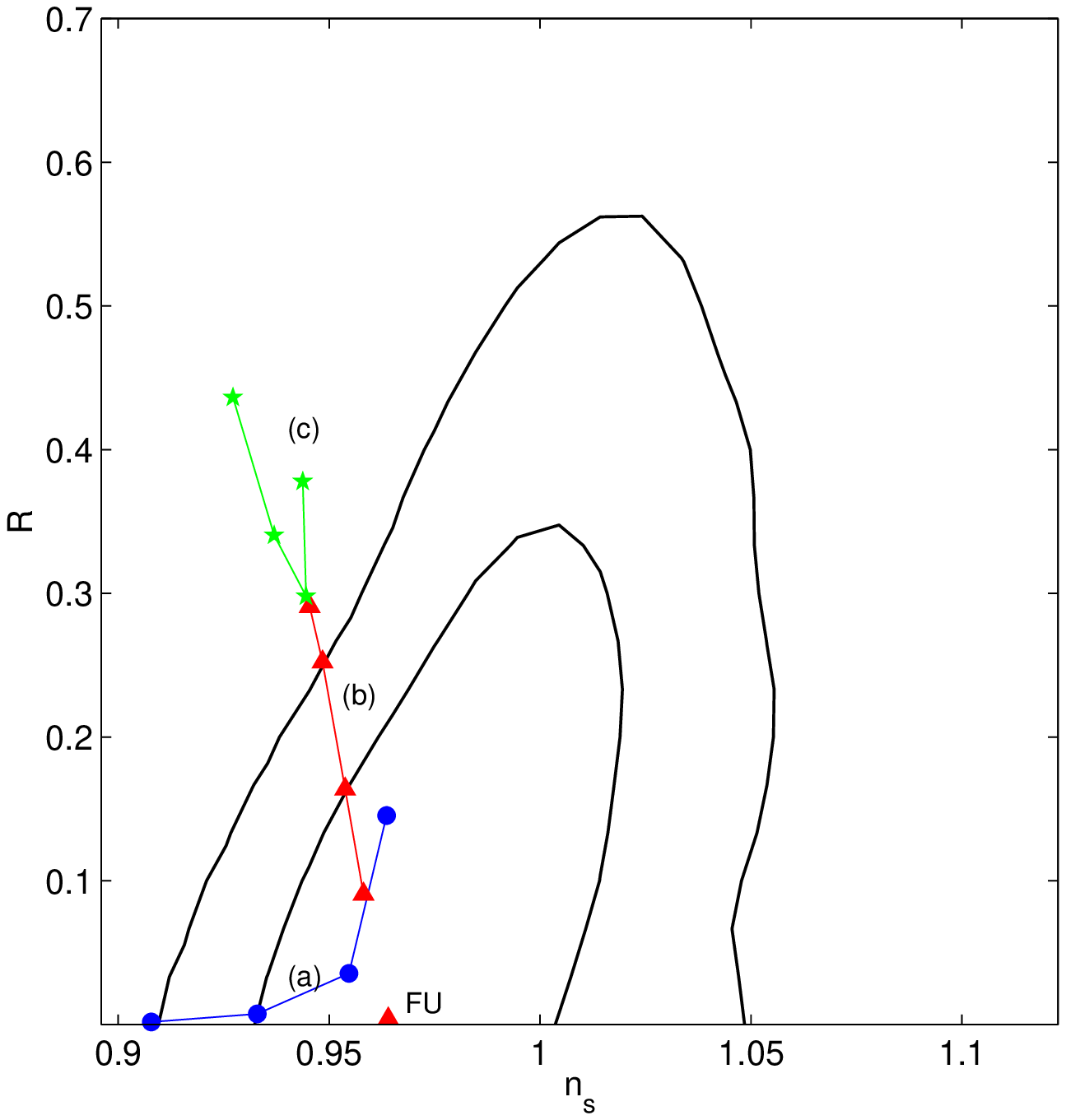}
\begin{figcaption}{nsR}{10cm}
2D posterior constraints in the $n_{\rm S}$-$R$ 
plane with the $1\sigma$ and $2\sigma$ 
contour bounds. We also show the theoretical 
predictions for (a) $p=2$, (b) $p=4$ and (c) $p=6$
with a fixed $e$-fold, $N=55$.
Each case corresponds to, from top to bottom, 
(a) $\xi=0, -0.003, -0.007, -0.011$ and (b)
$\xi=0, -0.0003, -0.0017, -0.005$. The point denoted by 
``FU'' is the Fakir and Unruh scenario with $|\xi| \gg 1$.
The plot (c) shows the cases of $\xi=0, -0.001, -0.0035$
from the left top to bottom, and another point corresponds to 
$\xi=-0.01$.
\end{figcaption}
\end{center}
\end{figure}

In Fig.\,\ref{nsR} we plot the 2D posterior constraints in the $n_{\rm S}$-$R$ 
plane and also show the $1\sigma$ and $2\sigma$ 
contour bounds.
In the previous subsection we obtained $n_{\rm S}$ and $R$ in terms of the function of $\xi$ by fixing the $e$-fold to be $N=55$.
Therefore one can constrain the strength of nonminimal couplings 
by plotting theoretical predictions of $n_{\rm S}$ and $R$
in the same figure. 

\subsubsection{Case of $p=2$}

In the case of the quadratic potential, the theoretical point is within
the $1\sigma$ contour bound for $\xi=0$, 
as is seen in Fig.\,\ref{nsR}.
Taking into account negative nonminimal couplings leads to the 
decrease of both $n_{\rm S}$ and $R$ (see Figs.~\ref{ns} and \ref{R}).
{}From Fig.\,\ref{nsR} we obtain the observational constraint on 
negative nonminimal couplings:
\begin{eqnarray}
& &\xi>-7.0 \times 10^{-3}~~~(1\sigma~{\rm bound})\,, \\
& &\xi>-1.1 \times 10^{-2}~~~(2\sigma~{\rm bound})\,.
\end{eqnarray}
If $\xi$ is less than of order $10^{-2}$, 
the curvature of the potential around 
$\phi=\phi_{\rm M}$ is too steep to generate a nearly scale-invariant 
spectrum.

\subsubsection{Case of $p=4$}

It is now well known that the quartic potential is 
under a strong observational 
pressure in the minimally coupled 
case \cite{Peiris,Barger,Kinney,Leach,Tegmark}.
In fact the $\xi=0$ point is outside of the $2\sigma$ contour bound
in Fig.\,\ref{nsR}.
In the presence of negative nonminimal couplings, 
we have the increase of 
$n_{\rm S}$ and the decrease of $R$, which is favoured observationally.
Figure \ref{nsR} indicates that $\xi$ is constrained to be 
\begin{eqnarray}
& &\xi<-1.7 \times 10^{-3}~~~(1\sigma~{\rm bound})\,, \\
& &\xi<-3.0 \times 10^{-4}~~~(2\sigma~{\rm bound})\,.
\end{eqnarray}
Thus nonminimal couplings of order $\xi=-10^{-3}$ make it possible 
to generate observational preferred power spectra.
In Fig.\,\ref{nsR} we also plot the theoretical point in the limit of 
$|\xi| \to \infty$ (denoted by ``FU''). 
This corresponds to the values given in Eq.~(\ref{nRp=4limit})
which is deep inside the $1\sigma$ contour bound.
Thus the Fakir and Unruh scenario with $|\xi| \gg 1$
is favoured observationally relative to the minimally coupled case.
In addition this scenario can relax the fine tuning problem
of the coupling constant $c$ \cite{FU}.

\subsubsection{Case of $p=6$}

The $p=6$ case is far away from the $2\sigma$ bound for $\xi=0$.
Negative nonminimal couplings lead to the increase of $n_{\rm S}$
and the decrease of $R$ when $|\xi|$ is much smaller than unity.
However this behaviour is altered with the growth of $|\xi|$, as we showed
in the previous section. The tensor-to-scalar ratio $R$ is minimum around 
$\xi=-3.5 \times 10^{-3}$, whose point is outside 
of the $2\sigma$ contour
bound. Since one has  the decrease of $n_{\rm S}$
and the increase of $R$ for $\xi<-3.5 \times 10^{-3}$, this regime is also away from the $2\sigma$ bound. 
Therefore the $p=6$ case is disfavoured observationally 
even in the presence of nonminimal couplings.
This situation does not change for $p>6$, since the theoretical points tend 
to be away from the observationally allowed region for larger $p$. 
In fact we numerically checked that the $p=8$ case is outside of the 
$3\sigma$ bound for any values of $\xi$.

\section{Conclusions and discussions}

In this work we studied cosmological perturbations in generalized
gravity theories based on the action (\ref{lag}).
We showed that curvature perturbations in the Jordan frame 
coincide with those in the Einstein frame.
Since tensor perturbations are also invariant  
under a conformal transformation, the inflationary observables
($n_{\rm S}$, $n_{\rm T}$, $R$ and $A_{\rm S}$) are the same
in both frames. This property indicates that the same likelihood
can be employed as for the standard Einstein gravity.
This is similar to what happens for the Randall-Sundrum II
braneworld scenario in which the degeneracy of the consistency 
relation does not explicitly give rise to the signature of the 
braneworld \cite{Huey}, although the 
constraints of model parameters in terms of underlying potentials
are different \cite{TL}.
Remarkably the consistency relation 
(\ref{slow}) holds even for the generalized action (\ref{lag}) 
that includes dilaton gravity, JBD theory, and a nonminimally 
coupled scalar field.

We then apply our general formalism to the nonminimally coupled
inflaton field with potential (\ref{potential}).
Our main aim is to place strong constraints on the strength of 
nonminimal couplings using the latest observational data
including WMAP, the 2dF and SDSS galaxy redshift surveys.
We focused on the case of the negative nonminimal couplings, since 
the positive coupling $\xi$ was already 
severely constrained from the requirement of a sufficient 
inflation \cite{FM} 
($\xi$ is at least smaller than $10^{-3}$ and is 
even much smaller depending on 
the initial condition of the inflaton).

For the quadratic potential ($p=2$), inclusion of 
negative nonminimal couplings leads to the decrease of the spectral 
index $n_{\rm S}$ and the tensor-to-scalar ratio $R$
(see Figs.\,\ref{ns} and \ref{R}).
While the minimally coupled case ($\xi=0$) is within the 
$1\sigma$ contour bound, the theoretical points of larger $|\xi|$ 
tend to be away from the observational bounds 
due to the departure from the scale-invariance of the 
spectral index (see Fig.\,\ref{nsR}).
We found the constraints $\xi>-7.0 \times 10^{-3}$
at the $1\sigma$ level and 
$\xi>-1.1 \times 10^{-2}$ at the $2\sigma$ level.

The quartic potential ($p=4$) suffers from a strong observational 
pressure for $\xi=0$, since the $\xi=0$ case is outside of the 
$2\sigma$ contour bound. However this situation is easily 
improved by taking into account negative nonminimal 
couplings, as seen in Fig.\,\ref{nsR}.
The strength of the coupling is constrained to be $\xi<-1.7
\times 10^{-3}$ at the $1\sigma$ level and 
$\xi<-3.0 \times 10^{-4}$ at the $2\sigma$ level.
Note that the Fakir and Unruh scenario with large negative $\xi$
($|\xi| \gg 1$) is deep inside the $1\sigma$ bound, thus preferred
observationally.
We also found that the $p \ge 6$ cases are outside of the $2\sigma$
bound even in the presence of negative nonminimal 
couplings (see Fig.\,\ref{nsR}).

While we mainly concentrated on slow-roll inflation, the formula
(\ref{tilns}), (\ref{nTein}) and (\ref{ratioep}) can be used 
in more general theories if the terms $\dot{\epsilon}$, $\dot{\delta}$ and $\dot{\beta}$
vanish. Actually this happens for the dilaton gravity 
($F=e^{-\phi}$ and $\omega=-e^{-\phi}$) with 
an exponential potential. Let us consider a negative exponential
potential, $\hV=-V_0 \exp(-\sqrt{2/\alpha}\,\hat{\phi})$, which 
appears in the Ekpyrotic scenario \cite{ekpyr}.
Note that this potential is the one in the Einstein frame and
the dilaton $\phi$ is related with the separation of two parallel branes
$\hat{\phi}$ through the relation 
$\phi=-\sqrt{2}\hat{\phi}$ \cite{TBF}. In this case 
the background evolution is characterized by 
\begin{eqnarray}
\hat{H}=\frac{\alpha}{\hat{t}}\,,~~~~
\frac{{\rm d}\hat{\phi}}{{\rm d}\hat{t}}
=-\frac{\sqrt{2\alpha}}{\hat{t}}\,,
\label{HEEin}
\end{eqnarray}
in the Einstein frame and 
\begin{eqnarray}
H=-\frac{\sqrt{\alpha}}{t}\,,~~~~
\phi=-\frac{2\sqrt{\alpha}}{1-\sqrt{\alpha}}
{\rm ln}\,\left[-(1-\sqrt{\alpha})t\right]\,,
\label{Hstring}
\end{eqnarray}
in the string frame \cite{TBF}.
Then we have $\epsilon=1/\sqrt{\alpha}$, 
$\hat{\epsilon}=-1/\alpha$,
$\delta_{\rm S}=\delta_{\rm T}=\beta=-1/(1-\sqrt{\alpha})$
and $\hat{\delta}_{\rm S}=\hat{\delta}_{\rm T}=0$, which 
are all constant. 
Therefore (\ref{tilns}), (\ref{nTein}) and (\ref{ratioep}) 
can be {\em exactly} employed in spite of 
the fact that the background evolution
is not slow-roll. The spectral indices $n_{\rm S}$, $n_{\rm T}$
and the tensor-to-scalar ratio $R$ are invariant under a
conformal transformation, and simply given by 
\begin{eqnarray}
\hat{n}_{\rm S}=n_{\rm S}=1+\frac{2}{1-\alpha}\,,~~~
\hat{n}_{\rm T}=n_{\rm T}=\frac{2}{1-\alpha}\,,~~~
\hat{R}=R=\frac{16}{\alpha}\,,
\label{nsntR}
\end{eqnarray}
which are highly blue-tilted spectra for 
$0<\alpha \ll 1$ \cite{bounceper}.
Note that these are the spectra generated during the contracting 
phase and may be affected by the physics around the 
bounce \cite{CDC}. We simply presented this example in order
to show the validity of the formula
(\ref{tilns}), (\ref{nTein}) and (\ref{ratioep}) rather than working 
on the detailed evolution of perturbations in the bouncing cosmology.
We note that the formula
(\ref{tilns}), (\ref{nTein}) and (\ref{ratioep}) are automatically 
valid in slow-roll inflation, since the variation of the terms
$\epsilon$, $\delta$ and $\beta$ are negligibly small.

There exist other generalized gravity theories where the function 
$F$ in the action (\ref{lag}) depends upon not only $\phi$ but also 
the Ricci scalar $R$. 
The $R^2$ inflationary scenario proposed 
by Starobinsky \cite{R2} belongs to this class, in which 
the spectrum of density perturbations was derived in 
Refs.\,\cite{Sta,Kofman,Hwang01}. In particular Hwang and 
Noh \cite{Hwang01} showed that the spectra of both 
scalar and tensor perturbations are invariant under a conformal 
transformation as in our action (\ref{lag}).
These facts imply that the degeneracy of the consistency relation 
persists in a wide variety of gravity theories 
including the higher-curvature gravity theory
and the Randall-Sundrum II braneworld.

\section*{ACKNOWLEDGMENTS}
We are grateful to Sam Leach for providing the latest SDSS
code and for his kind help in implementing the likelihood 
analysis. We also thank Antony Lewis and David Parkinson 
for their support in technical details of numerics.
The research of S.T. is financially supported 
from JSPS (No.\,04942). 
S.T. thanks to all members in IUCAA for their warm 
hospitality during which this work was completed 
and especially to Rita Sinha for her kind support
in numerics. 


\end{document}